\documentclass[journal]{vgtc}                

\usepackage{mathptmx}
\usepackage{mathrsfs}
\usepackage{graphicx}
\usepackage{times}
\usepackage{amsmath}
\usepackage{amssymb}
\usepackage{amsfonts}
\usepackage{algorithm}
\usepackage{subfigure}
\usepackage[american]{babel}
\usepackage{color, colortbl}
\usepackage{paralist}
\usepackage{url}
\usepackage{booktabs}
\usepackage{multirow}
\usepackage{changepage}
\usepackage{moresize}
\usepackage[numbers,sort]{natbib}

\usepackage[bookmarks,backref=true,linkcolor=black]{hyperref} 
\hypersetup{
  pdfauthor = {},
  pdftitle = {},
  pdfsubject = {},
  pdfkeywords = {},
  colorlinks=true,
  linkcolor= black,
  citecolor= black,
  pageanchor=true,
  urlcolor = black,
  plainpages = false,
  linktocpage
}

\onlineid{121}

\vgtccategory{Research}
\vgtcpapertype{Design Study}

\vgtcinsertpkg

\title{SkyLens: Visual Analysis of Skyline on Multi-dimensional Data}

\author{Xun Zhao, Yanhong Wu, Weiwei Cui, Xinnan Du, Yuan Chen, Yong Wang, Dik Lun Lee, and Huamin Qu}
\authorfooter{
\item{X. Zhao, X. Du, Y. Chen, Y. Wang, D. Lee, and H. Qu are with the
    Hong Kong University of Science and Technology. E-mail: \{xzhaoag, xduac, ychencd, ywangct, dlee, huamin\}@ust.hk}
\item{Y. Wu is with the IBM T.J. Watson Research Center, Yorktown Height, NY and is the corresponding author. E-mail: yanhongwu@outlook.com}
\item{W. Cui is with Microsoft Research Asia. E-mail: weiweicu@microsoft.com}
}

\shortauthortitle{Biv \MakeLowercase{\textit{et al.}}: Global Illumination for Fun and Profit}

\abstract{
Skyline queries have wide-ranging applications in fields that involve multi-criteria decision making, including tourism, retail industry, and human resources.
By automatically removing incompetent candidates, skyline queries allow users to focus on a subset of superior data items (i.e., the skyline), thus reducing the decision-making overhead.
However, users are still required to interpret and compare these superior items manually before making a successful choice.
This task is challenging because of two issues.
First, people usually have fuzzy, unstable, and inconsistent preferences when presented with multiple candidates.
Second, skyline queries do not reveal the reasons for the superiority of certain skyline points in a multi-dimensional space.
To address these issues, we propose SkyLens, a visual analytic system aiming at revealing the superiority of skyline points from different perspectives and at different scales to aid users in their decision making.
\changed{Two scenarios demonstrate the usefulness of SkyLens on two datasets with a dozen of attributes.}
A qualitative study is also conducted to show that users can efficiently accomplish skyline understanding and comparison tasks with SkyLens.
}


\CCScatlist{ 
 \CCScat{K.6.1}{Management of Computing and Information Systems}%
{Project and People Management}{Life Cycle};
 \CCScat{K.7.m}{The Computing Profession}{Miscellaneous}{Ethics}
}

  \teaser{
 \centering
  \includegraphics[width = \linewidth]{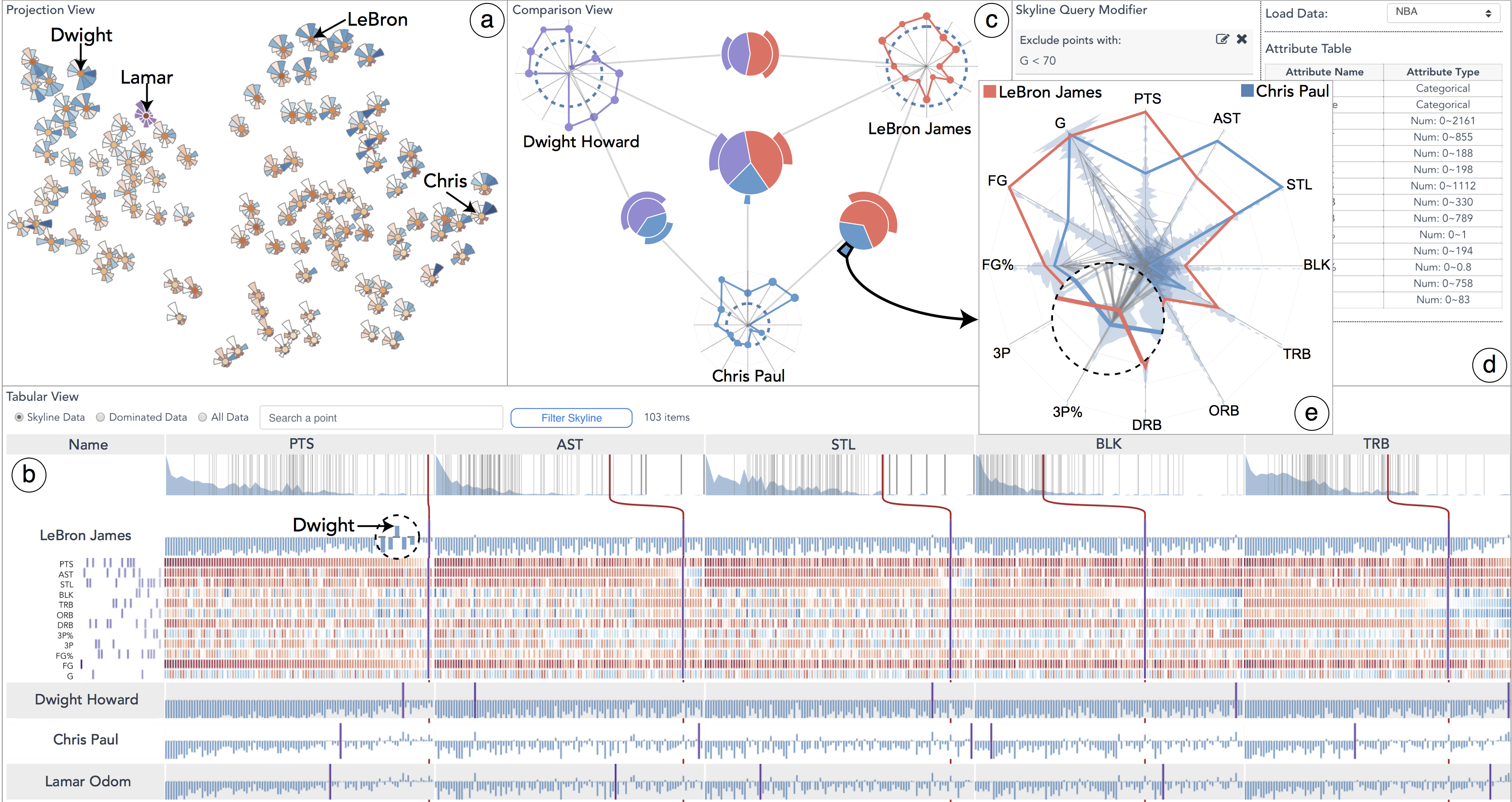}
    \vspace{-4mm}
	\caption{Analyzing the skyline of NBA statistics using SkyLens: (a) a Projection View showing an overview of clusters and outliers; (b) a Tabular View depicting the attributes of four skyline players and reveals the factors making a player in skyline; (c) a Comparison View examining the differences between skyline players from the attribute and domination perspectives; (d) a Control Panel for refining skyline queries; (e) a pop-up window showing a detailed comparison between LeBron James and Chris Paul.}
  	\label{fig:teaser}
    \vspace{-1mm}
  }

\begin{document}


\definecolor{darkred}{RGB}{168,37,17}
\definecolor{darkblue}{RGB}{23,55,119}
\definecolor{highblue}{RGB}{20, 20, 180}
\definecolor{lightblue}{RGB}{183,210,237}
\definecolor{gray}{RGB}{100,100,100}
\definecolor{lightgray}{RGB}{230,230,230}
\definecolor{rainforest}{RGB}{3,101,100}
\definecolor{darkpurple}{RGB}{66,8,91}
\definecolor{orange}{RGB}{242, 101, 34}
\definecolor{black}{RGB}{0, 0, 0}
\newcommand{\outline}[1]{\textbf{\textcolor{darkblue}{[#1]}}}
\newcommand{\secround}[1]{\textcolor{black}{#1}}
\newcommand{\changed}[1]{\textcolor{black}{#1}}
\newcommand{\delete}[1]{\textcolor{orange}{#1}}
\newcommand{\note}[2]{\textcolor{blue}{#1 [#2]}}
\newcommand{\todo}[1]{\textcolor{red}{[#1]}}
\newcommand{\needrefine}[1]{\textcolor{darkred}{#1}}
\newcommand{\typos}[1]{\textcolor{black}{#1}}
\newcommand{\notsure}[1]{\textcolor{gray}{[#1]}}
\newcommand{\weiwei}[1]{\textcolor{blue}{#1}}
\newcommand{\ww}[1]{\textcolor{red}{$\leftarrow$#1$\rightarrow$}}
\newcommand{\etal}{et~al.\ }
\newcommand{\eg}{e.g.}
\newcommand{\ie}{i.e.}
\newcommand{\graphmargin}{\vspace{-.5mm}}

\newcommand{\highlight}[2]{\textcolor{red}{#1 [#2]}}


\firstsection{Introduction}
\label{sec:intro}
\maketitle
Given a multi-dimensional dataset, skyline queries automatically prune the dataset to a \changed{\textit{subset of superior points}} that are not dominated by others; this \changed{\textit{subset}} is referred to as skyline~\cite{borzsony2001skyline}.
Skyline queries are important in various fields that involve multi-criteria decision making, such as tourism \cite{papadias2005progressive}, retail industry \cite{chan2006finding}, and human resources~\cite{pei2005catching}, in which users need to compare candidates in a multi-dimensional dataset and make a decision.
For example, a tourist needs to select a vacation destination from a list of cities on the basis of several attributes, including cost, climate, quality of service, and safety.
If city A is less desirable in every attribute than city B (i.e., A is dominated by B), then skyline queries will remove A from the candidate list because whenever A is preferred, B is always a better choice under any circumstances.
Thus, skyline queries may significantly reduce the number of candidates for the tourist without affecting his/her final choice.

However, skyline queries only solve half of the problem, because users still have to select the most ideal item manually based on their personal preference.
In the aforementioned example, travel agents generally cannot decide which city is the best for the tourist.
Instead, the agents can only present all superior cities with their pros and cons to the tourist to decide.
To make a successful decision, users need to completely understand the semantics of the skyline and compare various skyline points, which is rather difficult, especially when the data are multi-dimensional and the points do not dominate one another.
In addition, the size of skyline may become excessively large when many dimensions are considered~\cite{gao2010finding}.
Previous research suggests that a large number of candidates or attributes could create emotional stress to users and act as powerful barriers to rational decisions~\cite{rachmawati2006preference}.
Thus, an assistant tool for systematically organizing, interpreting, and comparing skyline points is necessary to help users in decision making.

One major challenge in this task is interpreting skyline, that is, understanding what merits make a point in skyline~\cite{pei2005catching}.
A skyline point may excel because of a single attribute, a subspace of attributes, several subspaces, or all attributes.
All the subspaces that make a point in the skyline are called decisive subspaces, which may vary from point to point and are important to understand the skyline data~\cite{pei2005catching}.
In the aforementioned example, a city may be in the skyline only because it has the lowest cost.
However, another city may not perform the best in any single attribute, but no other city can beat it in both service quality and cost.
Individual tourists may have different preferences; thus, exploring the decisive subspaces of each skyline point can assist tourists in identifying a subset of interests further.
However, such information is hidden because skyline queries only capture the set of superior points without revealing the reasons leading to their superiority.

Another crucial issue is the comparison of multiple skyline points.
Skyline queries focus on the comparison within individual attributes, and thus lack the capability to relate different attributes to one another.
Normally, the attributes are combined by calculating the weighted sum of all the attribute values, but how to assign the exact weight to each attribute to represent user's preference is still a challenge~\cite{pajer2017weightlifter}.
Previous research suggests that people often have fuzzy, unstable, and inconsistent preferences when provided with multiple options~\cite{dhar1995new}.
\changed{In addition, people can only focus and construct rankings among a few options, neglecting other possibilities.}
Thus, a systematic and unbiased method to compare skyline points is needed.
Apart from attribute values, users also need to consider other unique characteristics of skyline points, such as the number of points dominated by certain skyline points~\cite{gao2010finding} and the decisive subspaces~\cite{pei2005catching}. As a result, the comparison becomes more complex.


To tackle these challenges, we develop SkyLens, a visual analytic system that aims to reveal the superiority of skyline points from different perspectives and facilitate decision-making processes.
To solve the first challenge, we propose a novel tabular design that summarizes the attribute-wise rankings and differences between individual skyline points. This design allows users to inspect the reasons for the superiority of certain skyline points.
The decisive subspaces of individual skyline points, which are important elements for skyline interpretation (Sec.~\ref{sec:designgoals}), are also illustrated to facilitate the decision-making process.
For the second challenge, two linked visualizations are integrated into SkyLens to help users compare skyline points at two scales: a macro-level for identifying clusters and outliers in the entire skyline and a micro-level for examining the differences in a small set of skyline points.
Moreover, SkyLens supports the analysis of the domination relations in the skyline to provide an additional but necessary perspective for skyline comparison.
The main contributions of this work are summarized as follows:
\begin{compactenum}
    \item An interactive visualization system, named SkyLens, to help users organize, interpret, and compare skyline points from different perspectives and at different scales.
    \item A novel tabular design utilizes diverging in-cell bar charts and contiguous matrices to provide in-depth details of individual skyline points and to facilitate skyline interpretation.
    \item Two use scenarios with real datasets and a qualitative user study to demonstrate the effectiveness and usefulness of SkyLens.
\end{compactenum}
 
\section{Background}
\label{sec:background}

\begin{figure}[!tb]
\centering
\includegraphics[width=0.8\linewidth]{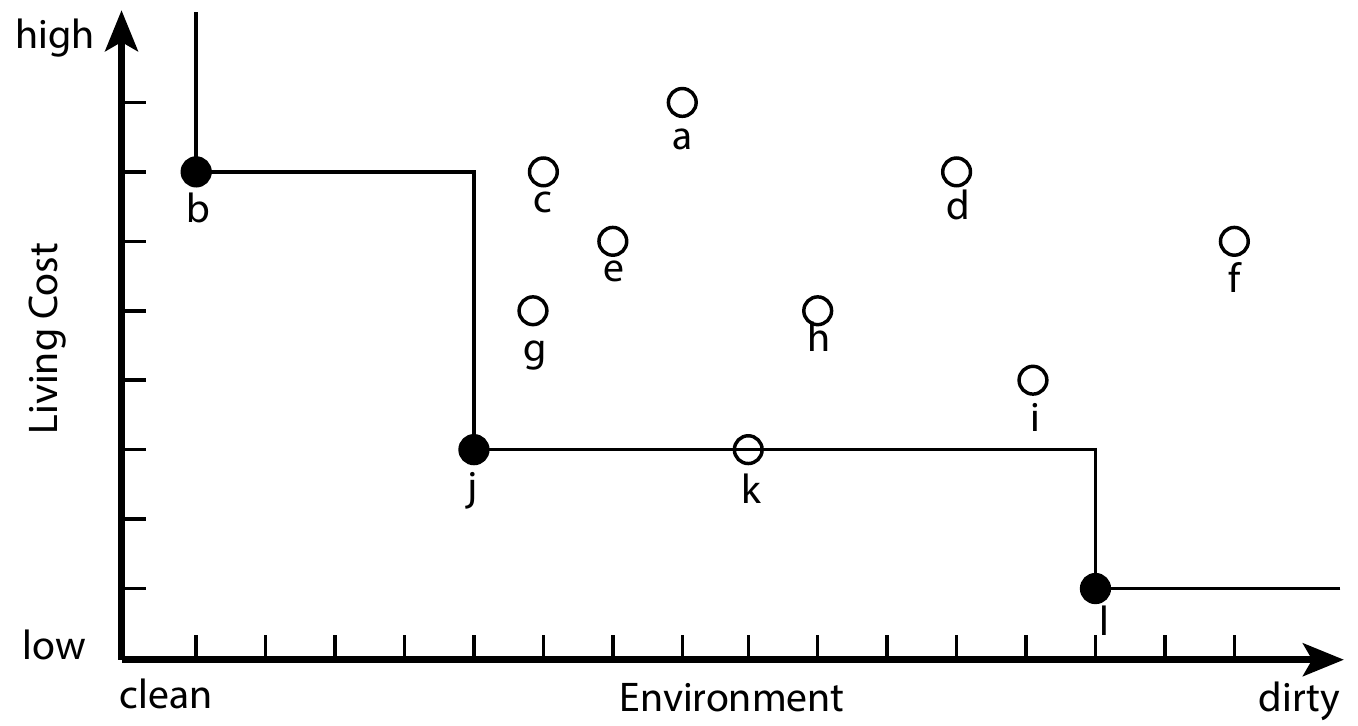}
\vspace{-3mm}
\caption{Example of a travel destination dataset with two attributes: living cost and environment. The solid black points $b$, $j$, and $i$ form the skyline of this dataset.}
\label{fig:skyline_example}
\vspace{-6mm}
\end{figure}

\changed{
	Assume a tourist wants to find a city that has both a clean environment and a low living cost.
	Fig.~\ref{fig:skyline_example} shows all possible candidate cities as a scatter plot, where each point represents a city.
	Some comparisons are obvious.
	For example, city $b$ dominates city $a$, as $b$ is cleaner and has a lower living cost.
	However, it is not obvious for cities $b$, $j$, and $i$, since they are not dominated by any other cities.
	Thus, these three cities form the skyline of the dataset.
	Once the skyline is extracted, the tourist can then safely neglect the rest cities, since the final choice is always from the skyline, disregarding his/her personal preference over these two attributes.
}


Formally, given an $m$-dimensional space $D = (d_1, d_2, \ldots, d_m)$, we denote $P = \{p_1, p_2, \ldots, p_n\}$ as a set of $n$ data points on space D.
For a point $p \in P$, it can be represented as $p = (p^1, p^2, ..., p^m)$ where $p^i \in \mathbb{Q} (1 \leq i \leq m) $ denotes the value on dimension $d_i$.
For each dimension $d_i$, assume that there exists a total order relationship on the domain values, either `$>$' or `$<$'.
Without loss of generality, we consider `$>$' (i.e., higher values are more preferred) in the following definitions.

\textbf{Dominance:} For any two points $p, q \in P$, $p$ is said to dominate $q$, denoted by $p \succ q$, if and only if $(i)$ $p$ is as good as or better than $q$ in all dimensions and $(ii)$ at least better than $q$ in one dimension,
i.e., $(i)~\forall~d_i \in D, p^i \ge q^i$ and $ (ii)~\exists~d_j \in D,~p^j > q^j$ where $1 \leq i, j \leq m$.

\textbf{Skyline point:}
A point $p \in P$ is a skyline point if and only if $p$ is not dominated by any $q \in P - \{p\}$,
i.e., $ \nexists q \in P - \{p\}$, $q \succ p$.

\textbf{Skyline:}
The skyline $A$ of $P$ is the set of skyline points in dataset $P$ on space $D$.

\textbf{Dominated point:}
A point $p$ is a dominated point if and only if there exists a point $q~(\neq p) \in P$ dominates $p$,
i.e., $\exists q \in P - \{p\}$, $q \succ p$.

\textbf{Dominating score:}
Suppose $A$ is the skyline of $P$, for a skyline point $p \in A$, the dominating score of $p$ is the number of points dominated by this point.
The dominating score can be denoted as $\phi(p)$, in which $\phi(p) = |\{q \in P - A |~p \succ q\}|$

\textbf{Subspace:}
Each non-empty subset $D'$ of $D$ is referred to as a subspace,
i.e., $ D' \subseteq D~\&~D' \neq \emptyset$

\textbf{Subspace skyline:}
For a point $p$ in space $D$, the projection of $p$ in subspace $D'\subseteq D$, denoted by $p^{D'}$, is in the subspace skyline if and only if $p^{D'}$ is not dominated by any other points $q^{D'}$.


\textbf{Decisive subspace:}
For a point $p \in P$ that is a skyline point in space $D$, if a subspace $B$ is decisive, if and only if that for any subspace $B'$ such that $B \subseteq B' \subseteq D$, $p^{B'}$ is in the corresponding subspace skyline.

\section{Related Work}
\label{sec:rel}

\subsection{Skyline Query}
\label{sec:rel-skyline-query}
Skyline queries can automatically extract superior points from a multi-dimensional dataset, which is very useful in multi-criteria decision making applications.
In research on skyline queries, a large number of studies aim to address two main drawbacks of skyline queries aside from developing algorithms that can more effectively process and accelerate skyline queries~\cite{kossmann2002shooting,papadias2005progressive,morse2007efficient}.
As the dimensionality increases, the size of skyline becomes large, causing failure for the skyline in providing interesting insights to users.
The other problem is that skyline queries do not incorporate user's preferences for different attributes.

Considerable effort has been devoted to generating a representative skyline from the entire skyline to reduce the skyline size in a high-dimensional space and to increase the discriminating power of skyline queries~\cite{papadias2003optimal,lin2007selecting,yiu2007efficient}.
They chose the $k$ most \textit{interesting} points by a metric of interestingness from the full skyline.
One category contributes to identifying a small subset of skyline that best summarizes the entire skyline.
For example, Tao et al.~\cite{tao2009distance} proposed the concept of distance-based representative skyline, in which the skyline points are clustered and the center point of each cluster is used as the representative subset of skyline.
Another group of studies quantifies interestingness numerically and ranks the skyline objects according to the numerical metric~\cite{gao2010finding,chan2006finding,nanongkai2010regret}.
Chan et al.~\cite{chan2006high} proposed \textit{skyline frequency}, which is defined as the number of subspaces that a point is in the skyline, to rank the skyline points and then return the top-$k$ frequent skyline points.
Although these metrics can reflect some aspects of the skyline point, they can not represent the specific needs of every end user.
Users may not be aware of these underlying metrics as well.
The actually interesting items could be missed when only the top $k$ items identified by user-oblivious metrics are provided to users.

Another drawback of skyline queries is that it treats all attributes as equally important.
In reality, however, users may not be interested in the skyline of full space (all attributes are considered) but rather in a subset of attributes~\cite{tao2006subsky,pei2007computing,pei2005catching}.
Several studies have attempted to integrate user preferences for attributes into the skyline queries and then reduce the skyline points of real interest.
Lee et al.~\cite{lee2009personalized} proposed an algorithm named \textit{Telescope}, which identifies personalized skyline points by considering both user-specific preferences over attributes and retrieval size.
Mindolin and Chomicki~\cite{mindolin2009discovering} proposed the \textit{p-skylines} framework, which augments skyline with the concept of attribute importance.
They developed a method to mine the relative importance of attributes from user-selected tuples of superior and inferior examples, which they have incorporated into the skyline queries.
These studies show that incorporating users' preferences for attributes can assist in filtering interesting points, but only few of them involved the real users.
By contrast, our system allows users to directly select attributes of interest and helps them select the most desirable point.

\subsection{Visualization for Multi-Criteria Decision Making}
\label{sec:rel-mcdmvis}

Ranking is one of the most popular methods for decision making, and ranking-based techniques can be applied to various applications such as billboard location selection~\cite{liu2017smartadp}, path finding~\cite{partl2016pathfinder}, and lighting design~\cite{sorger2016litevis}.
When weight is set to each attribute and the weighted attribute values are aggregated, multi-dimensional data points can be converted into scalar values and ranked according to these values.
Many visualization techniques have been proposed to help users dynamically adjust attribute weights and explore the relationships between weights and rankings.
For example, ValueCharts~\cite{carenini2004valuecharts} uses stacked bar charts to represent attribute weights and provides an immediate ranking feedback based on the aggregated weight values.
Lineup~\cite{gratzl2013lineup} further highlights the ranking changes after weight adjustment and allows users to compare multiple rankings and the corresponding weight settings simultaneously.
To analyze the relationships between ranking changes and weight modification, Weightlifter~\cite{pajer2017weightlifter} proposes the concept of weight space, which represents the ranges of the potential weights that guarantee a certain data point being ranked at the top positions.
However, though these methods allow users to set different weights to attributes iteratively, the process of finding a set of accurate weights that represent a specific user preference remains tedious and ineffective.
In fact, user preferences are often fuzzy and difficult to capture by a single weight.
Moreover, the preference of a user for an attribute may even be influenced by other attribute values.
For example, a tourist may not select a travel destination when the safety index of this place is excessively low regardless of how beautiful its environment is.
The reality complicates the weight-adjustment process and requires a heavy mental overhead from users.

Another popular approach to assist decision making is skyline queries.
Without requiring additional input from users, skyline queries can significantly reduce the size of candidates that users need to consider.
To facilitate skyline understanding, several multi-dimensional data visualization techniques have been leveraged.
For example, Lotov et al.~\cite{lotov2013interactive} visualized the bivariate relationships of skyline using the scatter-plot matrices, in which the points in each scatter-plot are colored according to their values in the third attribute.
All the other attributes are assigned to a certain value and users can use a slider to adjust values and explore skylines.
Andrienko et al.~\cite{andrienko2003building} improved this approach by adding a bar chart to show the distribution of differences between a specific skyline point and other points.
To support analyzing skyline in all dimensions simultaneously, some studies utilize Parallel Coordinates to visualize skyline~\cite{bagajewicz2003pareto}.
However, these approaches suffer from the visual clutter problem when the number of skyline points is large, a problem that prevents users from gaining insights into the skyline.
Projection-based methods have also been considered to help users explore skyline points.
For example, Shahar et al.~\cite{chen2013self} combined glyphs and the Self-Organizing Map (SOM)~\cite{kohonen1998self} to present skyline points and their affiliation to different attributes.
Although this solution provides an overview of skyline, projection and orientation errors could occur when more than three dimensions are considered in SOM~\cite{kohonen1998self}.
These errors may also mislead users in skyline interpretation without providing detailed skyline information.

In summary, these visualization techniques mainly focus on representing an overview of the whole skyline, which is not sufficient to support the decision making process that includes
exploring the whole skyline, narrowing down to a small subset, examining a few points in detail, and finally making a decision.

\section{Design Goals}
\label{sec:designgoals}
We have distilled the following design goals based on a thorough literature review of $50$ papers we collected from the database field and our interviews with two domain experts who work on skyline algorithms.
Further details are provided in our supplementary materials.

\textbf{G1: Explore the entire skyline from different perspectives and at different scales.}
Although skyline techniques can automatically exclude points that are dominated by superior ones, users still need to select their favorites themselves.
To make a quick and confident selection, users need to explore and understand the entire skyline from different perspectives and at different scales.
On the basis of our review, the goal is the first and most important, with $35$ papers focusing on this objective from different angles.
Our first design goal is critical for skyline analysis for two reasons.
First, the number of skyline points is often large, which hinders users from gaining insights into skyline~\cite{yiu2007efficient}.
Although a number of previous studies (Sec.~\ref{sec:rel-skyline-query}) aim at providing different criteria to rank the skyline points or identifying a representative subset of skyline points, these criteria cannot fully represent the requirements and preferences of users.
Thus, it is necessary to follow Ben Shneiderman's Visualization Mantra~\cite{shneiderman1996eyes} and enable the dynamic exploration of skyline at different scales.
Second, the comparison between skyline points can also be complex in a high-dimensional space~\cite{lee2007approaching}.
For example, when comparing skyline points, users may not only want to consider the values of each attribute but also to explore the value distribution of other points~\cite{lee2007approaching}.
Furthermore, when deciding if a skyline point is unique for some specific requirements, users need to ascertain the number of points dominated by that particular skyline point and determine whether these points are dominated by other skyline points~\cite{gao2010finding}.
Therefore, the visualization system needs to support skyline analysis from different perspectives, including the attribute-related information and the domination relations.

\textbf{G2: Understand the superiority of skyline points.}
Aside from generating the superior skyline points from the entire dataset, users also need to know on what combinations of factors a skyline point dominates other points~\cite{pei2006towards,magnani2013skyview}.
Users can easily focus on the points of interest rather than on the entire set of skyline points by gaining this insightful information about skyline.
The reasons that make a point in skyline can be observed from the relative ranking of the point in each attribute, its differences with other skyline points, and its decisive subspaces~\cite{pei2005catching}.
From the relative ranking in each attribute, users can infer in which attributes a specific skyline point is superior to others.
With finer granularity, users can examine the reasons a skyline point is not dominated by other points from the pair-wise difference between attribute values.
When the relative rankings cannot provide enough information, the decisive subspaces can be exploited to understand on what combinations of attributes the skyline point is superior.
These insights help users better understand how the skyline points differ from one another and facilitate decision making.

\textbf{G3: Compare skyline points and highlight their differences.}
Users always need to compare multiple skyline points before a successful selection in multi-criteria decision making scenarios.
This task not only includes an overall browsing of the entire skyline~\cite{balke2005approaching}, but also a detailed comparison of a few skyline candidates\cite{valkanas2013skydiver}.
The attribute statistical information, such as the relative rankings of skyline points and the value distribution in each attribute, is helpful when raw attribute values cannot provide users with sufficient knowledge to make decisions~\cite{lee2007approaching}.
For example, the attribute value distribution is useful when users want to examine whether a designated candidate is strong enough in certain attributes and when users do not have prior knowledge about the data.
Apart from examining the attribute values and attribute statistics, the domination relation, i.e., the relation of the point sets that are dominated by different skyline points, are also important when examining multiple skyline points~\cite{gao2010finding}.
From the dominating score and domination relation, users can inspect the specific data distribution behind skyline~\cite{gao2010finding}, and select the appropriate skyline points that best match their domain requirements.


\textbf{G4: Support an interactive exploration and refinement of skyline.}
User preferences are dynamic during their data-exploration process~\cite{dhar1995new}; thus, users should be provided with a convenient mode in which they can refine the skyline algorithms by removing certain points, constraining the range of attribute values, or excluding non-essential attributes~\cite{mahmoud2015strong, balke2007user}.
Furthermore, as user's understanding of the data deepens with data exploration, they may tend to be more interested in certain attributes or data ranges~\cite{yuan2005efficient,lee2012interactive}.
Thus, allowing users to select attributes of interests and highlighting those points that act as the subspace skyline of these attributes is essential.
A rich set of interactions, such as linking and brushing, filtering, and searching, should also be supported to facilitate the aforementioned requirements.

\section{Analytical Tasks}
\label{sec:analyticaltasks}
To fulfill the aforementioned design goals, we have extracted the following analytical tasks.

\textbf{T1: Encode multi-dimensional attributes and statistics.}
Showing the attribute values is insufficient for multi-dimensional skyline analysis.
The relative ranking of skyline points in each attribute should also been shown because raw attribute values could be misleading.
Furthermore, when users have no prior knowledge about the data, they may need to examine the value distribution in this attribute for decision making.
Thus, our system should encode not only multi-dimensional attributes but also the attribute statistics of skyline ({\textbf{G1, G3}).

\textbf{T2: Encode decisive subspaces of each skyline point.}
The decisive subspaces of a skyline point can provide users with a different perspective to examine the reasons a point is in the skyline (\textbf{G1, G2}).
According to the decisive subspace definition, the attributes in decisive subspaces guarantee that the corresponding point is in the full-space skyline.
Therefore, the decisive subspaces help reveal the outstanding merits of a skyline point, especially when the relative rankings of the points in each attribute are too close to illustrate attribute differences.

\textbf{T3: Highlight the differences between multiple skyline points.}
Highlighting the differences between skyline points is useful not only when comparing different skyline points (\textbf{G3}), but also when inspecting the reasons for the superiority of a point in the skyline (\textbf{G2}).
To compare the relative strengths of different skyline points across all the attributes, the system should first summarize the skyline point differences in all dimensions as a whole.
Moreover, the system should highlight the differences between skyline points in each attribute on demand so that users can quickly identify how other points differ from a selected point.
The intersections of dominated points and the value distribution at these intersections also suggest the relationships and differences between skyline points.
The system should provide a clear and effective mode to represent these domination relations among the skyline candidates for a detailed comparison.

\textbf{T4: Identify the clusters and outliers of skyline points.}
To provide the whole picture of all the skyline points (\textbf{G1}), the visualization system should enable users to identify clusters and outliers as the initial step of data exploration.
For example, when looking at the skyline of NBA statistical data, the players can be categorized into several groups, such as good attackers, adept defenders, or astute passers.
Users may have interests in one of the clusters and conduct further data exploration and analysis on this cluster.

\textbf{T5: Analyze the domination relations between skyline points.}
To provide a different perspective in addition to attribute-related information for comparing different skyline points (\textbf{G1, G3}), the system should allow users to analyze the domination relations among multiple skyline points.
This task includes illustrating both the dominating score and the differences between the dominated points of the selected skyline points.
Users may also have some prior knowledge on the data and want to know whether points that can dominate a specific data item exist.
For example, a tourist may want to know superior travel destinations compared with a visited place that satisfied him/her.
Finding those skyline points that dominate a designated candidate is useful for users in multi-criteria decision making scenarios.

\textbf{T6: Support refining skyline queries.}
During data exploration, users may want to exclude certain attributes or data items so that the skyline queries better match their requirements.
Fresh candidates can also appear in the refined skyline after excluding the undesired points from the skyline query.
Supporting a convenient skyline query refinement, such as setting the value range or removing certain attributes, can provide users an efficient and effective skyline exploration experience (\textbf{G4}).
This feature also helps control the skyline size within a manageable range and avoid the visual clutter problem.

\textbf{T7: Support filtering skyline results.}
From the skyline, users might opt to focus on a highly interesting subset or on a few candidates.
For example, when selecting a travel destination, users may only be interested in the places that have a moderate climate.
Furthermore, users may select their own attributes of interest and only keep the subspace skyline of these attributes for consideration.
Thus, the system should support skyline filtering by brushing certain value ranges and generating subspace skylines (\textbf{G4}).

\section{SkyLens Design}
\label{sec:visualdesign}

\changed{Motivated by the above analytical tasks, we design SkyLens to allow users to explore and compare skyline points at different scales and from different perspectives.
Our prototype\footnote{\textit{{http://vis.cse.ust.hk/skylens}}} is implemented \changed{using} Flask~\cite{flask}, VueJS~\cite{vuejs}, and D3~\cite{d3}.}

\changed{The system consists of a data analysis module and a visual analysis module. 
In the data analysis module, we unify the raw data to ensure higher values are better (Sec.~\ref{sec:background}) and then compute skyline.}
The visual analysis module incorporates three major views: 1) the \textit{Projection View} (Fig.~\ref{fig:teaser}a) that provides an overview of the entire skyline to identify clusters and outliers; 2) the \textit{Tabular View} (Fig.~\ref{fig:teaser}b) that summarizes the attribute-wise rankings and differences between skyline points, thereby allowing users to understand what combination of factors make a point in skyline, and 3) the \textit{Comparison View} (Fig.~\ref{fig:teaser}c) that aims to compare a small set of skyline points from both the attribute and domination perspective in detail.
We also provide a \textit{Control Panel} (Fig.~\ref{fig:teaser}d) to help users load data and refine skyline queries such as removing some specific attributes or excluding certain points.
A set of interactions is also provided to help users explore and refine skyline freely by filtering, linking, and brushing.


\subsection{Projection View}
\label{sec:projectionview}
The Projection View aims at providing an overview of skyline to allow users to discover clusters and outliers (\textbf{T4}).
In addition, we design skyline glyphs to encode detailed attribute values of each point and help users compare different skyline points (\textbf{T1}).


\textbf{Projection layout.} The skyline points are projected onto a 2D space, and their relative similarities are reflected through their placements to help users discover clusters and outliers.
Many dimension reduction techniques, such as MDS~\cite{kruskal1978multidimensional} and PCA~\cite{peason1901lines}, may be used for this purpose.
In our system, we adopt the t-distributed stochastic neighbor embedding (t-SNE) algorithm because t-SNE repels dissimilar points strongly to form more obvious clusters~\cite{maaten2008visualizing}.
Subsequently, we construct a similarity matrix based on the Euclidean distance between skyline points and then use t-SNE to project all skyline points onto a 2D space.
Thus, the skyline is visualized so that similar points are placed nearby while dissimilar points are placed faraway.

\begin{figure}[!tb]
\centering
\includegraphics[width=1.0\linewidth]{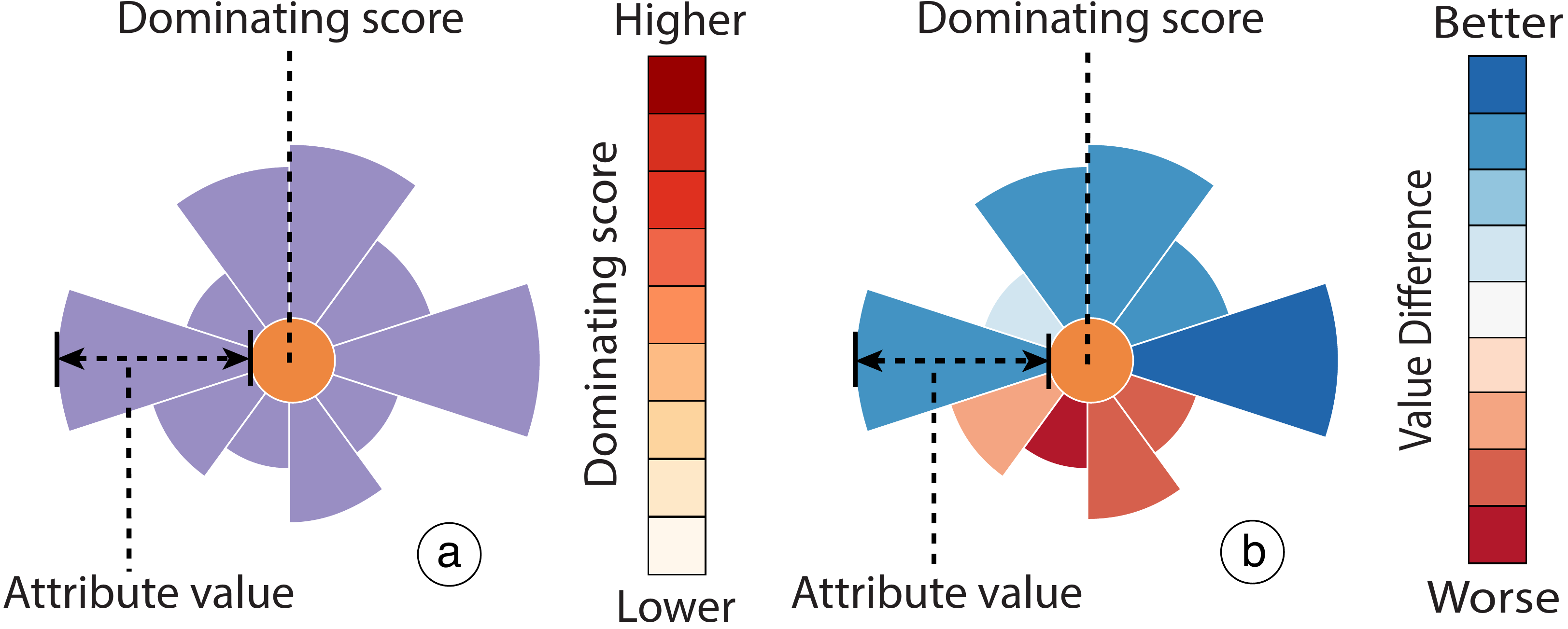}
\vspace{-6mm}
\caption{Skyline point glyphs in (a) the \textit{normal mode} and (b) the \textit{focus mode}. The inner circle color encodes the dominating score;
outer sector radiuses encode numerical values of attributes.}
\label{fig:glyph}
\vspace{-2mm}
\graphmargin
\end{figure}

\begin{figure}[!tb]
\centering
\includegraphics[width=1.0\linewidth]{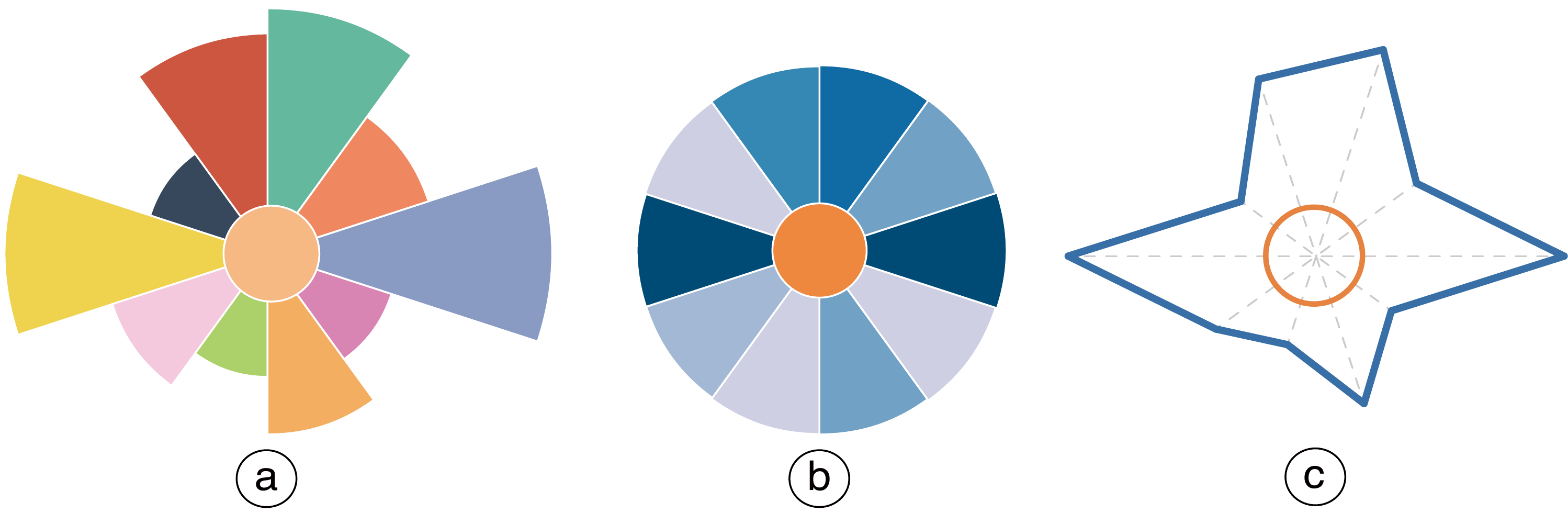}
\vspace{-5mm}
\caption{Three design alternatives for skyline point glyphs. All inner circles encode the dominating score. Attribute values are encoded differently: (a) using categorical colors to encode different attributes and using outer sector radiuses to encode numerical values; (b) using a sequential color scheme to encode numerical values; c) using a star glyph.}
\label{fig:glyph_alternatives}
\vspace{-5mm}
\end{figure}

\textbf{Skyline glyph.}
\changed{To better identify the differences between clusters and find representative skyline points, we further enhance the Projection View with glyphs in view of the effectiveness of glyphs in facilitating visual comparison and pattern recognition.}
Two fundamental metrics, namely, attribute values and dominating score, are used to differentiate skyline points and characterize clusters.
Accordingly, our glyph design is composed of two parts (Fig. \ref{fig:glyph}a): the inner circle and the outer sectors.
The inner circle color depicts the dominating score, where darker orange indicates a higher score.
The outer sectors represent the attribute values so that users can quickly identify skyline point clusters and outliers from the glyph shape.
To further assist in the comparison task of \textbf{T3}, we develop a \textit{focus mode} to enable users to obtain an intuitive overview of how a specific point differs from other skyline points.
When users select a glyph of interest, all the sectors of the other glyphs will be colored to highlight their differences from the selected one (Fig. \ref{fig:glyph}b).
For example, if the attribute value is higher than that of the selected glyph, the corresponding sector's color changes to blue.
This allows users to examine the differences between skyline points without changing the sector radius.

A potential drawback of the design is visual clutter, which is a common issue for many dimension reduction-based visualizations.
\changed{
To mitigate this problem, we first decrease glyph opacities so that individual glyphs can be observed.
When hovering over a glyph, the glyph will be enlarged and brought to the foreground.
In addition, we support panning and zooming to focus on a specific region of glyphs.}

\textbf{Glyph alternatives.}
During the glyph design process, we considered several design alternatives.
Our first design choice is between the circular design (e.g., radar charts) and the linear design (e.g., bar charts).
\changed{For the Projection View, we mainly focus on the overview of many glyphs.
Compared with circular designs, linear designs are more helpful when examining and comparing  different glyphs at a specific attribute.}
In addition, linear designs often require more space to achieve the same level of legibility as that of circular designs~\cite{mcguffin2010quantifying}.
Thus, a circular-based design is adopted in our system.

We also experiment on three design alternatives of circular design.
In these designs, the visual encoding of the inner circle is the same as that of our final design, which uses a sequential color scheme to show the dominating score.
However, these designs are all abandoned due to various reasons.
For example, our first alternative (Fig.~\ref{fig:glyph_alternatives}a) uses double encoding (i.e., categorical color and angle) to identify attributes.
\changed{However, categorical colors might be too distractive when there are many attributes.}
Our second alternative (Fig.~\ref{fig:glyph_alternatives}b) fixes the radius of the outer sectors and uses a divergent red-blue color to encode the numerical attribute value of each sector.
However, this design has two main drawbacks.
First, the color saturation is a less accurate visual channel compared to the length channel for encoding numerical values.
Second, for overlapping glyphs, the color blending may lead to a misinterpretation of values.
We also attempt using classic star glyphs to encode the attribute values (Fig.~\ref{fig:glyph_alternatives}c).
However, compared with our final design, the lines in the star glyphs are difficult to perceive when the color saturation is low and when the glyphs are small.
\subsection{Tabular View}
A major issue with skyline queries is that they only identify the skyline in the dataset without additional information.
Thus, we design the Tabular View to provide users with in-depth details about individual skyline points.
For example, users may want to know the difference between a specific skyline point and other skyline points to infer how good it is in the entire skyline (\textbf{T3}).
In addition, the decisive subspaces can help users understand how balanced a skyline point is (\textbf{T2}).
All these details are encoded in this view to provide users with insights into why and how a skyline point is superior, thereby facilitating the decision-making process (\textbf{T2, T3}).
To address the scalability issue, three interactions are also tightly integrated into this view to allow users to eliminate unsuitable skyline points rapidly (\textbf{T7}) and focus on the interesting subset of skyline points.

\textbf{Visual encoding.}
The Tabular View encodes the detailed information about each skyline point in an interactive tabular form (Fig. \ref{fig:tabular_view}). Attributes are encoded as columns in this view \changed{(e.g. \textit{Attr. I}, \textit{Attr. II}, and \textit{Attr. III} in Fig.~\ref{fig:tabular_view})}.
At the head of each column, an area plot shows the value distribution of all the data (Fig. \ref{fig:tabular_view}a), including both the skyline and the dominated points.
The $x$-axis represents the attribute value in an ascending order from left to right, while the $y$-axis represents the data density.
The skyline points are drawn as vertical gray lines on top of the area charts.
The combination of context area plots and foreground gray lines provides users with the distribution of the skyline lines and their places in the entire dataset to help them compare and evaluate the qualities of skyline points in terms of individual attributes.

Skyline points are represented as rows in the table \changed{(e.g.~\textit{ID A} and \textit{ID B} in Fig.~\ref{fig:tabular_view})}.
By default, all rows are displayed in the \textit{summary mode}, which summarizes the overall differences between skyline points.
Specifically, each table cell shows a diverging bar chart (Fig. \ref{fig:tabular_view}b).
\changed{We choose the linear bar chart design for focusing on the values of one single attribute in the data.}
Without lose of generality, we assume the table cell refers to skyline point $p_i \in \{p_1, p_2, \ldots, p_n\}$ and dimension $d_j\in\{d_1, d_2, \ldots, d_m\}$.
Accordingly, the cell has a total of $n$ bars, each representing a skyline point.
All bars are sorted (ascending) in accordance with their values at dimension $d_j$.
Among these bars, a special purple bar is placed to indicate the position of the skyline point $p_i$ in the sorted bars.
The height of each blue bar $s_{k}$, where $k\neq i$, represents the summarization of its differences from $p_i$ in all the other dimensions
(i.e., $\{d_l\}_{1\leq l\leq m}-\{d_j\}$).
Specifically,
$${\delta}_l(p_i,p_k)= (p_i^l-p_k^l) / \sqrt{\sum\nolimits_{i=1}^{n}{(p_i^l-\overline{p^l})^2/n}},$$
where $p_i^l$, $p_k^l$ are the values of $p_i$, $p_k$ at attribute $d_l$, and $\overline{p_i^l}$ is the mean value of attribute $d_l$.
Thus, the summary difference $\Delta(p_i,p_k) = \sum_{l=1}^{m}{\delta}_l(p_i,p_k)$, where ${l}\neq{j}$.
$\Delta(p_i,p_k)$ can be either positive or negative; thus, a horizontal dashed line is drawn in the middle of the table cell as a baseline.
The blue bars positioned above the baseline exhibit positive differences, whereas those below the baseline exhibit negative differences.
The summary mode is designed to help users compare skyline points from two aspects (\textbf{T3}).
First, users may select a table cell of interest to examine, and the position of the purple bar can give users a precise idea of the performance of the skyline point in terms of the attribute.
Then, the blue bars can further provide an overall idea of the performance of the skyline point in the other attributes.


Users can further click a row to expand it and examine its detailed comparison with other skyline points.
In this \textit{expansion mode}, we append a matrix below the diverging bar chart (Fig. \ref{fig:tabular_view}c).
In each small matrix, the columns represent the skyline points and are aligned with the bars in the diverging bar chart above the matrix.
The rows in the matrix represent the attributes and have the same order as the columns in the large table.
Each matrix cell is filled with a color to represent the difference between a specific skyline point (the matrix column $p_k$, where $k\neq i$) and the expanded one ($p_i$) in a specific attribute (the matrix row $d_l$), which is ${\delta}_l(p_k,p_i)$.


The decisive subspaces are also shown in the \textit{expansion mode}, specifically on the left side of the first detailed matrix (Fig. \ref{fig:tabular_view}d).
Each decisive subspace takes a vertical line, and each row represents a dimension.
Thus, if a dimension is involved in the decisive subspace, then a purple mark is placed in the corresponding space in the vertical line.
This visualization allows users to observe the number of decisive subspaces by counting the vertical lines.
The skyline points with numerous decisive subspaces are usually preferred because these skyline points are also strong in terms of different subspaces.
In addition, users can compare horizontally to identify which attributes are more involved in the decisive subspaces.
Thus, users may pay more attention to the skyline points that have attributes that are highly valued and involved in decisive subspaces because these attributes are often the merits of the corresponding skyline points.

\begin{figure}[!tb]
\centering
\includegraphics[width=0.95\linewidth]{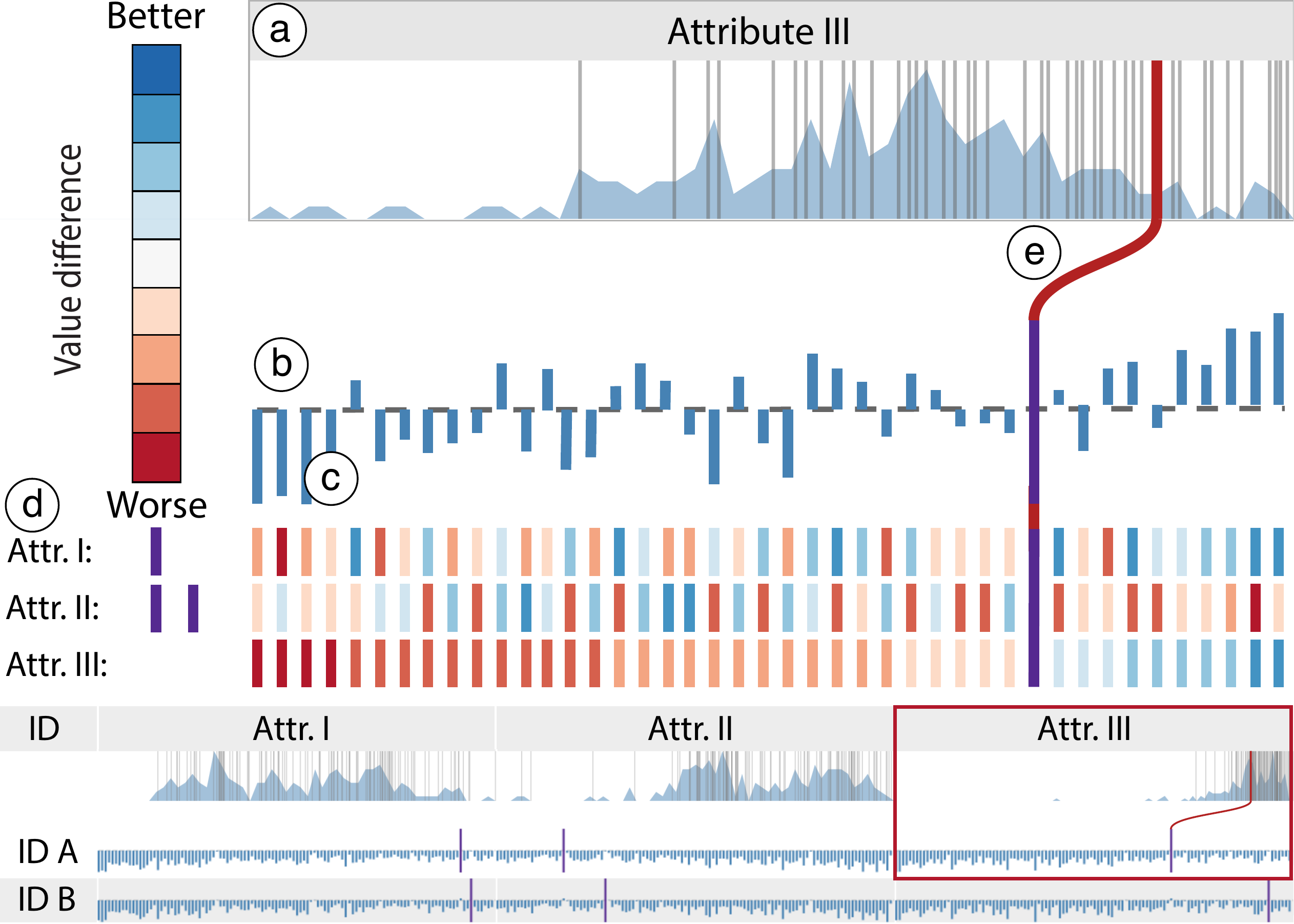}
\vspace{-2mm}
\caption{Visual encodings in the Tabular View: (a) the column header showing a specific attribute's value distribution; (b) the diverging bar chart depicting the point's relative ranking at this attribute and its overall differences with the other skyline points; (c) the \textit{expansion mode} showing the detailed comparisons between this point and other points at all attributes; (d) the bars representing the decisive subspaces of the point; and (e) the linking curve connecting the relative ranking and absolute value of the point at this attribute.}
\label{fig:tabular_view}
\vspace{-6mm}
\end{figure}

\textbf{Interactions.}
The Tabular View also supports the following user interactions to help users highlight attribute information (\textbf{T1}) or filter certain skyline points (\textbf{T7}):
\begin{compactitem}
\item{\textbf{Filtering}.}
\changed{SkyLens allows users to filter the skyline points by two modes: filtering a subspace of interest and filtering a subset of skyline points.
By clicking table headers, users can select certain attributes and highlight the skyline in the subspace of the selected attributes.}
SkyLens also supports users to brush on the area plot in each column header to indicate an acceptable region of attribute values.
If a skyline point does not meet the regional conditions, then the corresponding table row turns gray to reduce the number of interesting skyline points.
\item{\textbf{Linking.}}
When the cursor hovers over the row, several red lines appear to connect the purple bars to the corresponding gray lines in the table header (Fig. \ref{fig:tabular_view}e).
The divergent bars in a table cell only show the relative rankings of the skyline points in terms of the corresponding attribute, whereas the red linking lines can help users examine the raw values of all skyline points.
\item{\textbf{Searching.}}
\changed{Users with prior knowledge can search a specific point in the dataset using the search box at the top of the Tabular View.
If the point happens to be a skyline point, the corresponding row is highlighted.}
However, if the point is not in the skyline, SkyLens will highlight the skyline points that dominate the point.
\end{compactitem}

\subsection{Comparison View}
While the Projection View provides the whole picture of skyline and the Tabular View helps with reasoning about individual skyline points, the most important step is to thoroughly compare and examine the differences between a couple of candidates.
When users find desirable skyline points in the other views, they can click on the glyphs or rows to add them to this Comparison View for detailed comparison.
Apart from attribute values (\textbf{T3}), the number of dominated points and the value distribution of these dominated points are also important aspects to compare (\textbf{T5}).
Therefore, we design this view to allow users to closely investigate the differences among a few skyline points from these aspects.
Specifically, two types of visual elements are designed for this view: the radar charts for perceiving the attribute values of different skyline points and the domination glyphs to summarize and compare each skyline point's dominated points.

As shown in Fig. \ref{fig:comparison_view}, the added points are represented by the radar charts, which are arranged on a circle at uniformly distributed angles.
We adopt this circular layout to emphasize the comparison between skyline points by putting the domination glyphs in the center part of the view, thus letting users focus on the comparison quickly and directly.
Each domination glyph is connected to a number of radar charts and visually summarizes the differences between the connected skyline points.
If $n$ skyline points are selected, we enumerate all possible combinations (i.e., $\sum_{i=2}^n{{i}\choose{n}}$) and add a domination glyph for each of them.
Although the combination number grows exponentially, the scalability is not a big issue in our scenario, since we mainly focus on comparing a small number ($\leq 4$) of skyline points in this view.
A force-directed based layout is used to position the domination glyphs so that they can be arranged close to their linked radar charts.

\textbf{Radar charts}.
We use radar chart, a widely used multi-dimensional data visualization technique, to show the attribute values of selected skyline points (differentiated by categorical colors).
However, we enhance the traditional radar charts in several ways for our specific scenario as shown in Fig.~\ref{fig:comparison_view}a.
First, we draw circles on axes to encode the relative rankings of the skyline points in the corresponding dimensions.
Inside each polygon, we also draw a blue circle, whose radius represents the dominating score of the corresponding skyline point.
\changed{The design is consistent with the Projection View and is more space-efficient compared with affiliating additional indicators outside the radar chart.}
When hovering over a radar chart, a pop-up window will display to show more details (Fig.~\ref{fig:comparison_view}b).
On each axis, we draw the value distribution of the corresponding dimension, in which the values increase from the center along the axis and the width of the flow indicates the number of points.

\begin{figure}[!tb]
\centering
\includegraphics[width=1.0\linewidth]{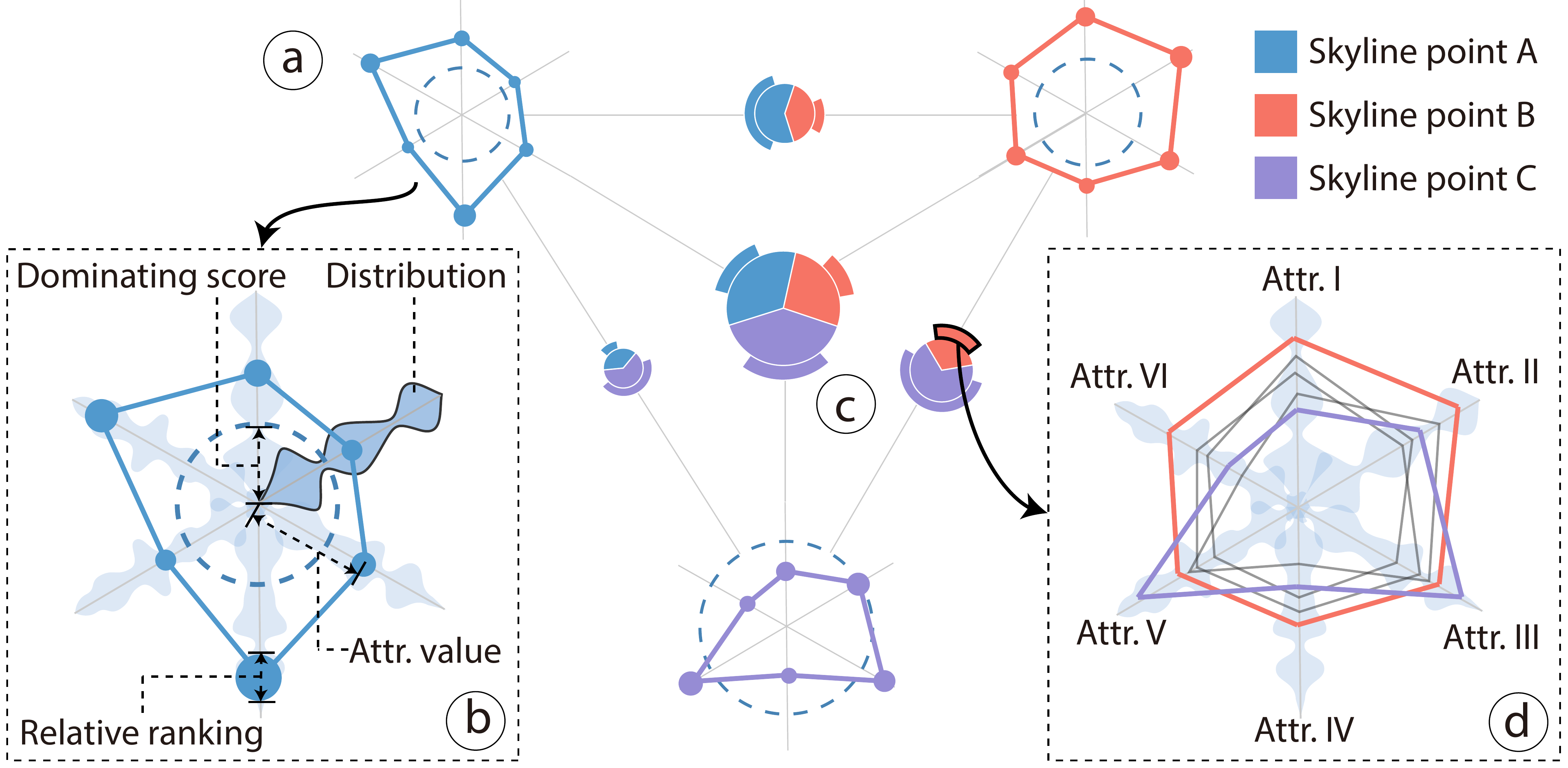}
\vspace{-5mm}
\caption{Visual elements of the Comparison View: (a) the radar chart that shows point A's attribute values and statistic information; (b) the visual encodings in the radar chart; (c) the domination glyph that summarizes the domination differences; and (d) a pop-up radar chart that illustrates the exclusive dominated points of point B.}
\label{fig:comparison_view}
\vspace{-6mm}
\end{figure}

\textbf{Domination glyph}.
The domination glyph is designed to summarize the differences between a small number of skyline points from the domination perspective.
\changed{We use a circular design based on the same consideration discussed in Sec.~\ref{sec:projectionview}.}
Similar to skyline glyphs, a domination glyph incorporates two parts (Fig.~\ref{fig:comparison_view}c).
An inner pie chart shows the dominating scores of the linked skyline points.
In addition, we also use the radius of the chart to encode the number of points that are dominated by at least one linked skyline point.
Surrounding the pie chart, arcs are displayed to represent the proportion of points that are exclusively dominated by the corresponding skyline points.

When hovering over an inner sector or an outer sector, a radar chart will also pop to show the comparison among these linked skyline points in detail.
The skyline points are represented by the thick colored lines and the dominated points are represented by the thin gray lines.
For example, the gray lines in Fig.~\ref{fig:comparison_view}d represent the points that are exclusively dominated by the orange skyline point.
This helps users identify the reasons these exclusively dominated points are not dominated by the other skyline points (i.e., in which attributes they have higher values than these skyline points).

\textbf{Alternative designs}.
Before adopting the current design, we also considered two alternatives (Fig.~\ref{fig:comparison_view_alter}), both of which have two parts: a central radar chart that compares the attribute values of skyline points and the outer rings that show the dominating scores.
In the first design (Fig.~\ref{fig:comparison_view_alter}a), the number of outer rings are equal to the number of selected skyline points, and each colored circle that is positioned on a ring represents a dominated point.
Thus, if several skyline points share a dominated point, the corresponding outer rings will each have a duplicate circle to represent this dominated point.
In the second design (Fig.~\ref{fig:comparison_view_alter}b), all the dominated points are illustrated on a single outer ring.
If a point is dominated by several skyline points, it will appear as a pie chart indicating the exact skyline points that dominate it.
\changed{A categorical color scheme is used to distinguish skyline points.}
\changed{However, both designs suffer from severe visual clutter due to the overlap of outer rings when the number of dominated points is large.}
For the above reasons, we abandon these two designs.

\subsection{Interactions}
\label{sec:interactions}
\changed{We developed a set of interactions to help users switch between the coordinated views.}
\changed{First, users can change the order of attributes in all the views by dragging the attribute rows in the Attribute Table (Fig.~\ref{fig:teaser}d).
In addition, when clicking a skyline glyph in the Projection View, not only will the skyline point be appended to the Comparison View, but the Tabular View will also automatically scroll to the row that represents this skyline point.}
Similarly, when hovering over a row in the Tabular View or hovering over a radar chart of the comparison view, the corresponding skyline glyph in the Projection View will be enlarged and moved to the foreground.
Besides, when brushing certain attribute ranges or calculating a subspace skyline, the results will also be highlighted in the Projection View.

\section{Evaluation}
\label{sec:evaluation}
\begin{figure}[!tb]
\centering
\includegraphics[width=1.0\linewidth]{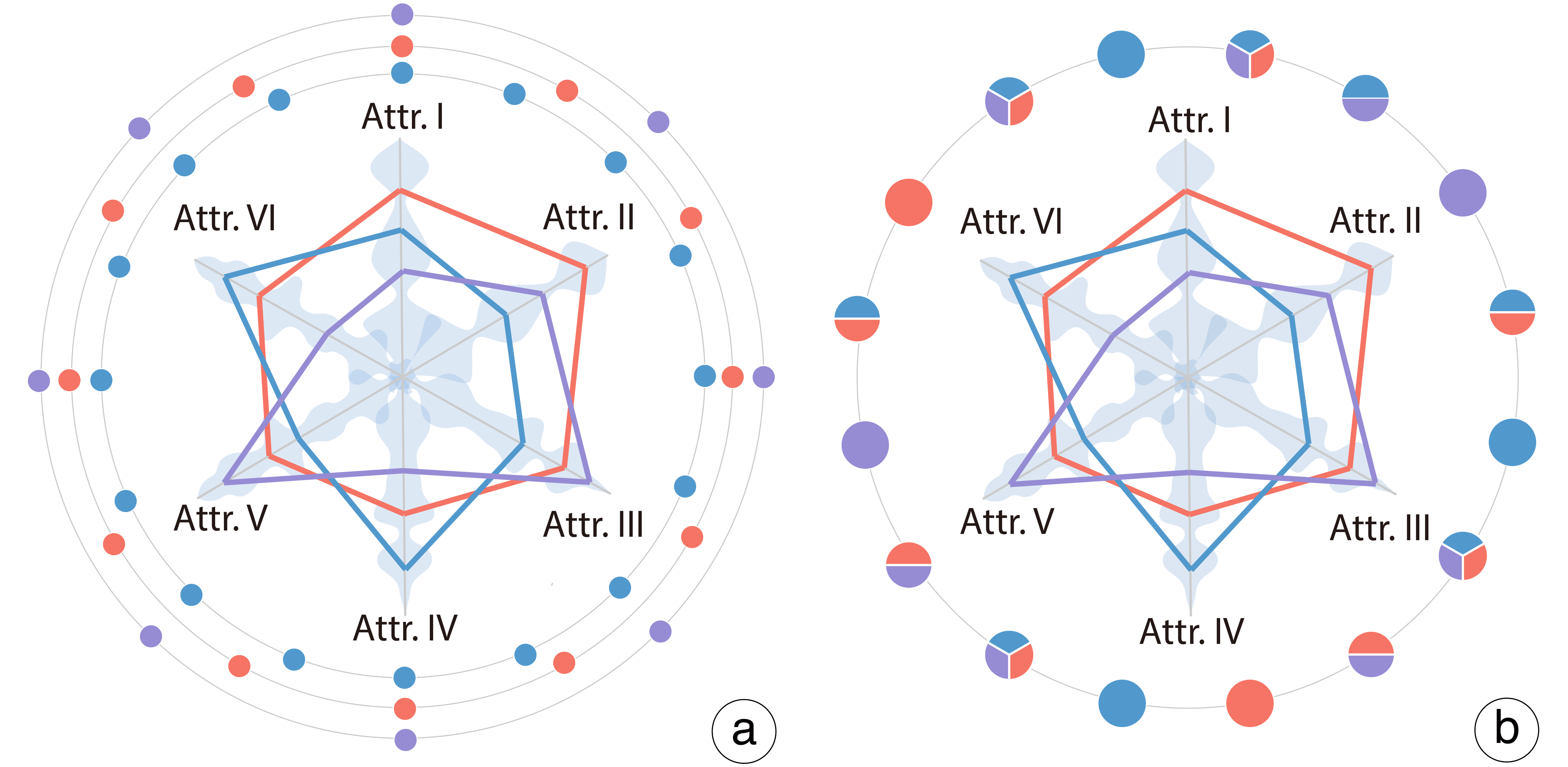}
\vspace{-5mm}
\caption{Two design alternatives for the Comparison View: (a) using small circles to represent the dominated points and outer rings to distinguish skyline points; (b) using pie charts to represent the dominated points and sector colors to distinguish skyline points.}
\label{fig:comparison_view_alter}
\vspace{-4mm}
\end{figure}

\begin{figure*}[ht]
\centering
\includegraphics[width=1.0\textwidth]{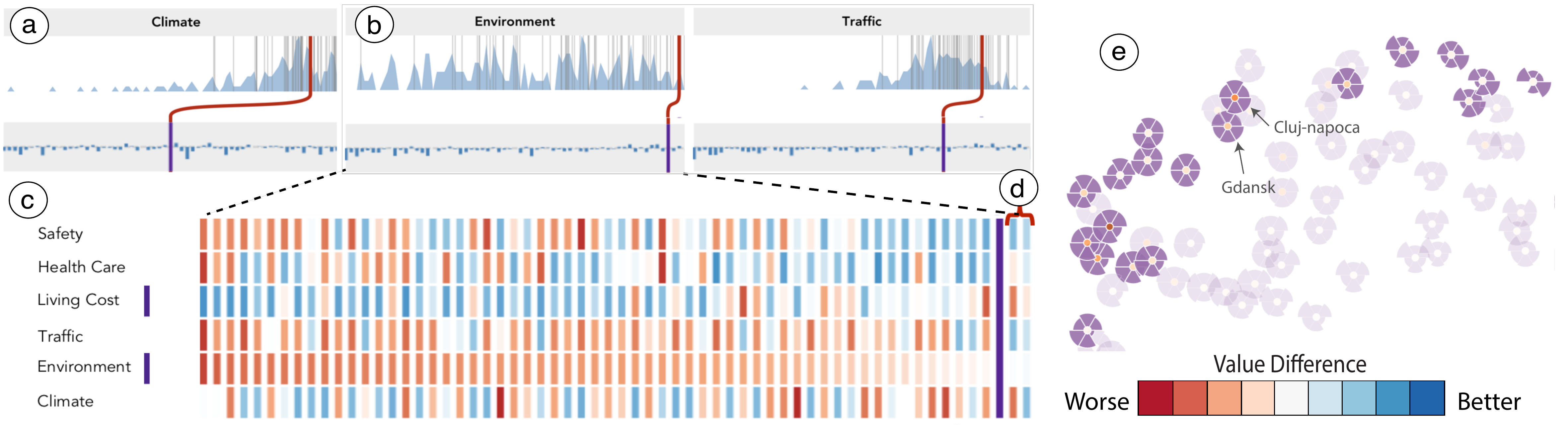}
\vspace{-5mm}
\caption{The Tabular View of Victoria: (a) the column header of \textit{Climate}, (b) the column headers of \textit{Environment} and \textit{Traffic}, (c) the decisive subspace, and (d) Wellington and Reykjavik that have higher value than Victoria in \textit{Environment}. (e) The Projection View that highlights the skyline of the subspace of \textit{Living Cost}, \textit{Traffic}, and \textit{Environment}.}
\label{fig:casestudy2}
\graphmargin
\vspace{-4mm}
\end{figure*}
\subsection{Usage Scenario I}
\label{sec:usagescenarioI}

The first usage scenario describes Alan, a journalist who wants to write an article about the most outstanding players of an NBA season.
He chooses not to rank the players because any ranking criteria can easily be criticized by NBA fans as different readers have different preferences.
Thus, he decides to use SkyLens to explore the specific merits of the most outstanding skyline players and to investigate the differences between them (\textbf{G3}).
He then loads the NBA 2010--11 regular season statistics, which include 452 players and 12 numerical attributes, such as \textit{Points Scored} (\textit{PTS}), \textit{Field Goals} (\textit{FG}), and so on \changed{(Fig.~\ref{fig:teaser})}.
Alan first looks at the Projection View and identifies several outliers that have rather small glyph sizes (\textbf{T4}).
After examination, he discovers that the outliers are players who only have high shooting percentages (\textit{FG\%} and \textit{3P\%}) and play only a few games.
Alan is not interested in these players, so he excludes the players who attend less than 70 games using the Control Panel (\textbf{T6}).
Then, he explores each skyline player's dominating score to find the player who outperforms the largest number of players in all attributes for this season.
By examining the inner circle colors, he finds Lamar Odom, who dominates 183 players in total.
Alan wants to further explore how other players are compared with Lamar Odom, he then double-clicks Lamar's glyph to switch the Projection View into \textit{focus mode}.

From the skyline glyph positions and the outer sector colors (Fig.~\ref{fig:teaser}a), he observes three major clusters of players (\textbf{T4}).
The players in the upper cluster (Fig.~\ref{fig:teaser}a) mostly have higher values in \textit{PTS} and \textit{FG} than Lamar, which indicates they are good scorers.
In this cluster, Alan identifies a skyline glyph with many large blue outer sectors (Fig.~\ref{fig:tabular_view}a), which represents LeBron James.
This means LeBron outperforms Lamar in almost half the attributes.
Next, he uses the Tabular View to examine the detailed information about LeBron.
From the positions of the purple bars in the row of LeBron (Fig.~\ref{fig:teaser}b), He observes that LeBron has high rankings in most of the attributes, which indicates that he is also a versatile player (\textbf{T1}).
In that row, Alan further observes that all the blue bars, which measure the overall differences between other skyline players and LeBron, are positioned beneath the baseline with one exception, Dwight Howard.
This suggests that Dwight Howard, who belongs to another cluster (Fig.~\ref{fig:teaser}a) in the Projection View, has an overall comparable performance with LeBron (\textbf{T3}).
To further compare these two players in detail, Alan opens the expansion mode of LeBron (Fig.~\ref{fig:teaser}b) to locate Dwight in the expanded matrix and observes that Dwight outperforms LeBron in the defense-related attributes, such as \textit{Total Rebounds} (\textit{TRB}) and \textit{Blocks} (\textit{BLK}).
To verify whether these defense-related attributes make Dwight in the skyline, Alan switches to the row of Dwight.
By examining the expanded matrix of Dwight, Alan identifies four defense-related attribute rows that are colored in red.
This indicates that no other player in the skyline has a better performance than Dwight in these defense-related attributes, which verifies his hypothesis.
In addition, he finds that many skyline players outperform him in the attribute \textit{Assists} (\textit{AST}).
He then checks if any of these players are located in the last cluster in the Projection View.
By highlighting the corresponding matrix bars, he identifies Chris Paul (Fig.~\ref{fig:teaser}a), a player who has the best performance in both \textit{Assists} (\textit{AST}) and Steals (\textit{STL}).

Since each of the three players (LeBron, Dwight, and Chris) can represents an individual cluster respectively in the Projection View, Alan decides to write a paragraph about how they dominate other players.
Thus, he adds these three players into the Comparison View.
From the central domination glyph (Fig.~\ref{fig:teaser}c) that summarizes their differences in dominating scores, Alan finds that LeBron and Dwight have almost the same number of players they dominate, while Chris only dominates half of the number of players and has very few exclusive players he dominates(\textbf{T5}).
When examining the other three pairwise domination glyphs, Alan observes that LeBron dominates almost all of the players that are dominated by Chris.
Considering Chris ranks much higher than LeBron in both \textit{AST} and \textit{STL}, it is strange for Chris to have a such limited number of players he exclusively dominates.
Alan investigates this phenomenon by hovering the cursor over the corresponding outer sector of the domination glyph. 
From the pop-up radar chart (Fig.~\ref{fig:teaser}e), he realizes that Chris also performs slightly better than LeBron in \textit{3P\%}, in addition to \textit{AST} and \textit{STL}.
In addition, the nine players that are exclusively dominated by Chris all have lower values in \textit{AST} and \textit{STL}, but higher values in \textit{3P\%} than LeBron.
To discover the underlying reason, Alan switches to the Tabular View and observes that few players have higher rankings than LeBron in either \textit{AST} or \textit{STL} from the distribution flow (Fig.~\ref{fig:teaser}b).
Thus, Alan understands why Chris does not dominate more players exclusively, although he performs extremely well at both \textit{AST} and \textit{STL}.



\subsection{Usage Scenario II}
\label{sec:usagescenarioII}

In the second usage scenario, we demonstrate how Lorraine, who is planning a one-month holiday, utilizes SkyLens to find a desirable city to visit.
Lorraine chooses to explore the Numbeo quality-of-life dataset~\cite{numbeodataset}, which includes 176 cities worldwide and 8 numerical attributes describing the overall living conditions of those cities.
Since Lorraine has little knowledge in how these attribute values are calculated, she decides to first use SkyLens to obtain outstanding cities and see their dominant attributes (\textbf{G2}).

Lorraine first excludes two traveler-irrelevant attributes by adding two filters in the Control Panel: the \textit{Purchasing Power} and the \textit{Housing Affordability}.
Since she has decided to spend her holiday outside Asia to experience a different culture, Lorraine also adds another filter to the \textit{Continent} attribute (\textbf{T6}) to exclude Asian cities.
Then, she regenerates skyline to obtain 62 candidate cities.

From the skyline cities, Lorraine identifies Victoria, a city in Canada, where she enjoyed the pleasant climate last summer.
She wants to further investigate what attributes make the city outstanding so that she can use it as a benchmark city.
Thus, she locates Victoria in the Tabular View and observes the purple lines that indicate its relative rankings in individual attributes.
Surprisingly, Lorraine finds that Victoria is just average in the attribute \textit{Climate} (Fig. \ref{fig:casestudy2}a).
However, by tracking the red curve that connects Victoria's relative ranking to the absolute value in the column header of \textit{Climate}, she realizes that most skyline cities perform well in \textit{Climate}.
In other words, most skyline cities have a moderate climate (\textbf{T1}).
Thus, the \textit{Climate} attribute is probably not a deciding factor for an ideal vacation destination.

On the other hand, Lorraine observes that Victoria has rather high relative rankings on \textit{Traffic} and \textit{Environment} (Fig.~\ref{fig:casestudy2}b).
Hence, she guesses these two attributes are what makes Victoria excel.
\changed{To verify this hypothesis, Lorraine switches to \textit{expansion mode} for the city and surprisingly discovers that the only decisive subspace of Victoria (\textbf{T2}) is (\textit{Living Cost}, \textit{Environment}) (Fig.~\ref{fig:casestudy2}c).}
Thus, at least one city is better than Victoria in both attributes.
To reveal these cities, Lorraine further examines the matrix in the \textit{expansion mode}.
She identifies that there are only two cities, Wellington and Reykjavik, ranking higher than Victoria in the \textit{Environment} attribute (Fig.~\ref{fig:casestudy2}d).
Nevertheless, these two cities also rank higher than Victoria in the \textit{Traffic}.
By further checking the matrix, Lorraine observes that Victoria only has higher values than Wellington and Reykjavik in the attribute of \textit{Living Cost}, which is consistent with Victoria's decisive subspace.

From her exploration of Victoria, Lorraine realizes that she wants to stay in a city that is good in \textit{Traffic} and \textit{Environment}, as well as having a reasonable value in \textit{Living Cost}.
In other words, she wants to find a city that is better than Victoria in the attribute of \textit{Living Cost}, while being close to Victoria, not necessarily better, in the attributes of \textit{Traffic} and \textit{Environment}.
After brushing the corresponding column headers, Lorraine sadly finds that no city satisfies all these requirements.
She decides to make a compromise and selects these three attributes (\textit{Living Cost}, \textit{Traffic}, and \textit{Environment}) as a subspace and highlight the cities in this subspace skyline (\textbf{T7}).

She switches to the Projection View (Fig.~\ref{fig:casestudy2}d) to examine the highlighted cities.
She observes that all the highlighted cities excel in at least two of the selected three attributes, while many cities have low values in a third attribute.
Since she does not want a city that has unacceptably low values in any attributes, only two cities, Gdansk and Cluj-napoca, are shortlisted.
She then switches to the Comparison View to compare these two points in detail, and finds that Gdansk has higher values in the attributes of \textit{Climate}, \textit{Traffic}, and \textit{Environment} than Cluj-napoca, which indeed satisfies her preferences.
Thus, she selects Gdansk as her travel destination.

\subsection{Qualitative User Study}
\label{sec:userstudy}

A formal comparative study with an existing skyline visualization system is not applicable because previous skyline visualization work mainly focuses the overview of skyline, which only covers a part of the tasks we list in Sec.~\ref{sec:analyticaltasks}.
Questions that involve interpreting and comparing skyline points require a complex examination from various aspects and cannot be simplified as yes/no questions.
Therefore, we choose to perform a qualitative study rather than a controlled quantitative experiment.
\changed{In addition to the qualitative study, we also conducted an informal comparison between our system and LineUp~\cite{gratzl2013lineup}, a ranking-based visual analytic tool to facilitate the decision making process.}

\textbf{Study design}.
We recruited 12 participants (3 females, aged 21 to 28 years (mean = 26.5, SD = 2.1)) with normal or corrected-to-normal vision.
All the participants were students in the computer science department of our local university.
Among them, 5 students had experience in information visualization and 3 students knew skyline queries.

We designed 10 tasks that covered all the important aspects in skyline analysis (Sec.~\ref{sec:analyticaltasks}) for the participants to perform.
The participants also needed to utilize all the views in SkyLens together to perform all the tasks successfully.
We also conducted several pilot studies to ensure that the study was appropriately designed.

The study began with a brief introduction of our system using the NBA dataset to help the participants get familiar with our system.
We also encouraged the users to freely explore our system after the introduction.
During this stage, we asked the participants to think aloud and ask questions if they encountered any problems.
To avoid memorization of data, we used the Numbeo quality-of-life dataset for the formal study.
For each task, we recorded the completion time and took notes of the feedback or problems raised by the participants for later analysis.
After the participants finished all the 10 tasks, we asked them to finish a questionnaire containing 19 questions about the usefulness and aesthetics of SkyLens.
Those questions were designed to evaluate our system in a 7-point Likert scale from strongly disagree (1) to strong agree (7).
In addition, we conducted an informal post-session interview with each participant to learn their opinion about our system in general.
During the interview, we also introduced LineUp to them and discussed with them about the differences between our system and LineUp in multi-criteria decision making scenarios.
\changed{We asked the participants to review all the tasks and suggest which of them can be performed using LineUp.}
On average, the entire study took approximately 40 minutes to finish.
The detailed task description, questionnaires, and study results can be found in our supplement materials.

\textbf{Results and discussion}.
All participants managed to complete the tasks in a short period of time (33.6s on average for each task).
However, task 8 took relatively longer time as it required the participants to manually search and add three cities into the Comparison View.
For the questionnaire results, most participants thought that it is easy to perform skyline analysis tasks using SkyLens (6.5).
They also reported that the system is visually pleasing in general (6.6), the interactions are easy in general (6.3), and the tool would be useful for many multi-criteria decision making scenarios (6.6).

In the post-session interviews, most participants appreciated the effectiveness and powerfulness of SkyLens as it can facilitate skyline understanding and enlighten the trade-off between attributes.
Specifically, they highlighted the usefulness of the Comparison View and the Tabular View.
Some participants commented that \textit{``The pie chart plus the outer sectors is indeed a smart design to help identify the domination differences between skyline points quickly.''}
While another participant added that \textit{``The Comparison View provides the flexibility to choose different combinations of players for comparison.''}
Some participants also appreciated the insights provided by the Tabular View.
Some participants reported that \textit{``The \textit{expansion mode} in the Tabular View is of great help to identify what combinations of attributes make a point outstanding, compared to the raw attribute values or rankings.''}

\changed{Apart from the positive feedback, the participants also suggest several improvements to our system.
For the Projection View, three participants, who used relatively long time to finish the Projection View's tasks, suggested to enlarge the default outer sector radius for the domination glyph to better compare the attribute values of different points.
Some participants also wanted to further examine a group of similar glyphs that locate together.
We adjusted the outer sector radius and enabled users to pan and zoom the Projection View after the interview.
For the Tabular View, a few participants who have no prior knowledge of skyline described that they needed some time to fully understand the visual encoding and the meaning of decisive subspaces.
This implies that the learning curve of SkyLens may be steep for people who have no experience in skyline analysis.
For the Comparison View, one participant reported that the sizes of some pie chart sectors are too small to select, thus we enlarged the minimum sector size accordingly.
}


As a comparison with LineUp, the participants reported that LineUp is a powerful tool and really easy to understand.
However, they all felt that LineUp can only support a small part of the tasks we focus (2.4), and SkyLens could cost less time when performing these tasks (6.6).
\changed{We believe this is because LineUp and SkyLens follow different approaches to decision making: ranking v.s. skyline analysis.}
\changed{For example, the participants identified that it was difficult to use LineUp to exclude the points that are dominated (i.e. worse in every aspects) by at least one point.
When comparing a few points in detail, the participants reported that they had to repeatedly perform multiple alignment interactions to determine the strong and weak attributes of different points using LineUp.}
In addition, the participants found that the weight adjustment process in LineUp is dubious and they usually did not know whether they had achieved the right weights to reflect their preferences.
When exploring the Numbeo dataset, one participant asked \textit{``Why Canberra always stays on the top? How can I change the weights to make other cities on the top of the list?''}
In summary, the participants reported that they would choose LineUp when they already have a good understanding about the dataset and a few trade-offs to consider.
Nevertheless, when they do not have enough prior knowledge of the dataset and want to carefully examine the data points from different perspectives, they preferred using SkyLens.
\changed{Thus, LineUp is desired when users know exactly their goals and SkyLen is preferred when users' requirements are vague and need detailed data comparison.
The two systems are complementary and are appropriate for different tasks in decision making scenarios.}

\section{Discussion}
One key issue in SkyLens is its scalability.
We adopted B\"orzs\"onyi et al.'s algorithm~\cite{borzsony2001skyline} in the system.
Although the algorithm has a high time complexity of $O(n^2m)$, where $n$ is point number and $m$ is the dimension number, it is sufficient for our experiment datasets.
However, efficient implementations~\cite{TanE01} may be adopted for larger datasets.
From the visualization perspective, 
too many skyline points may cause severe overlapping in the Projection View.
\changed{In our experiments, the Projection View can support more than a hundred skyline points with acceptable glyph overlaps.}
This issue can be further addressed by leveraging focus+context techniques.
In the Tabular View, the number of skyline points and attributes that can be displayed in the same window is also limited.
To address this issue, we design several interactions to help users exclude undesired skyline points and reorder attributes, so that they can place the most relevant information in the same view for exploration.
\changed{
The Projection View and the Comparison View can support the visualization of multi-dimensional data with about a dozen attributes.
For datasets with higher dimensions, though we enable users to visualize all the attribute values, it might be difficult to perceive the values due to the small sector angles.
Users can use the Control Panel of SkyLens to select attributes of their interests for further exploration.}
\changed{Besides, though the vanilla skyline algorithm only supports numerical attributes, some skyline variants also consider categorical attributes with a partial order.
In the future, We will enable users to define the order of categorical attributes; thus make SkyLens support categorical attributes.}

\changed{
Although SkyLens is designed to facilitate decision making, it can be easily extended to solve more general problems for multi-dimensional data exploration and analysis.
One example is multi-objective optimization, in which users need to identify a point $x$ in a multi-dimensional database to optimize $k$ objective functions.
SkyLens can facilitate this task by using these $k$ functions as data attributes; then calculate and visualize the skyline of the updated data.
In the future, we aim to embed an attribute editor to help users flexibly define and modify the data attributes for skyline analysis.
Given a multi-dimensional dataset, we can also build a network model where the nodes indicate a multi-dimensional point and the edges represent the domination relation between points.
By adopting network analysis methods, SkyLens can further support identifying clusters and outliers.
}
\section{Conclusion}
In this work, we propose SkyLens, a visual analytic system that assists users in exploring and comparing skyline from different perspectives and at different scales.
It comprises three major views: 1) the Projection View that presents the whole picture of skyline for identifying clusters and outliers; 2) the Tabular View that provides the detailed attribute information and the factors that make a point in skyline; 3) and the Comparison View that aims at comparing a small number of skyline points in detail from both the attribute value perspective and the domination perspective.
We also provide a rich set of interactions to help users interactively explore skyline.

In the future, we first plan to include nominal attribute analysis in SkyLens.
Many objects in multi-criteria decision making scenarios have nominal attributes; enabling users to dynamically edit their preferences on nominal attributes would extensively expand the application scope of SkyLens.
Furthermore, we want to investigate skyline visualization techniques that can support data with uncertain values, which is also a common scenario in many domains that requires skyline analysis.
At last, we aim to further explore how to track the temporal changes of skyline to assist temporal data analysis.
We hope this work will shed light on the future research on skyline visualization.
\section{Acknowledgments}
The authors wish to thank the anonymous reviewers for their valuable comments. This research was supported in part by HK RGC GRF 16208514 and 16241916.
\bibliographystyle{abbrv}
{\footnotesize
\bibliography{article}}

\end{document}